# EXACT SOLUTIONS: ANISOTROPIC STARS IN TERMS OF PRESSURE

**Naveen Bijalwan**[1]

Dev [6] discussed some exact solutions of anisotropic stars for special forms of TOV. Considering Bijalwan [1] ansatz for charged perfect fluids we present here some exact solutions of generalized TOV equation for anisotropic fluids by representing equation in terms of radial pressure. Consequently, radial pressure is found to be an invertible arbitrary function of $\omega(=c_1+c_2r^2)$, where $c_1$ and $c_2\,(\neq 0)$ are arbitrary constants, and $r$ is the radius of star, i.e. $p_r = p_r(\omega)$. We present a general solution for anisotropic fluid in terms for $\omega$. We list and discuss some old and new solutions which fall in this category. Consequently, we present solutions for generalized TOV considering generalized forms of anisotropy factor in terms of $\omega$. Also, we investigated solutions of generalized TOV with negative density gradient (NDG).

## 1. Introduction

In recent years a number of authors have studied solutions to the Einstein field equations corresponding to anisotropic matter where the radial component of the pressure differs from the angular component. The gravitational field is taken to be spherically symmetric and static since these solutions may be applied to relativistic stars. A number of researchers have examined how anisotropic matter affects the critical mass, critical surface red-shift and stability of highly compact bodies. These investigations are contained in the recent papers Dev et al. [6][7] , Chaisi et. al.[4][5], Maharaj et. al.[10], Mak et. al. [11][12], Sharma et. al. [16], Rao et. al. [13], Ivanov [9], Sharif [14], Cattoen [3], Hernández [8], among others. Some researchers have suggested that anisotropy may be important in understanding the gravitational behaviour of boson stars and the role of strange matter with densities higher than neutron stars. Mak et. al. [12] and Sharma et. al. [15] suggested that anisotropy is a crucial ingredient in the description of dense stars with strange matter.

Also, we presented solutions to generalized TOV equation by using methodology suggested by Bijalwan [1] and choosing a general equation of state which relates the radial pressure to energy density. Subsequently, we provide new solutions for NDG. Sharma et. al. [16] has demonstrated that the linear equation of state is consistent in the modeling of compact matter such as neutron stars and quasi-stellar objects.

## 2. Field Equations

Let us take the following spherically symmetric metric to describe the space-time of a charged fluid sphere

$$ds^2 = -e^{\lambda}dr^2 - r^2(d\theta^2 + \sin^2\theta d\phi^2) + e^{\nu}dt^2 . \tag{2.1}$$

are arbitrary functions.

[1]FreeLancer, c/o Sh. Rajkumar Bijalwan, Nirmal Baag, Part A, Pashulock, Virbhadra, Rishikesh, Dehradun-249202 (Uttarakhand), India. ahcirpma@rediffmail.com

Assuming the energy momentum tensor for an anisotropic star in the most general form

$$T_{ij} = diag(-\rho, p_r, p_\perp, p_\perp) \tag{2.2}$$

The Einstein field equations become

$$-\frac{\upsilon'}{r}e^{-\lambda} + \frac{(1-e^{-\lambda})}{r^2} = -\kappa p_r, \tag{2.3}$$

$$-\left[\frac{\upsilon''}{2} - \frac{\lambda'\upsilon'}{4} + \frac{\upsilon'^2}{4} + \frac{\upsilon' - \lambda'}{2r}\right]e^{-\lambda} = -\kappa p_\perp, \tag{2.4}$$

$$\frac{\lambda'}{r}e^{-\lambda} + \frac{(1-e^{-\lambda})}{r^2} = \kappa c^2 \rho, \tag{2.5}$$

where $\rho$ is the energy density and, $p_r$ and $p_\perp$ are the radial and tangential pressures respectively. Primes denote differentiation with respect to $r$.

Let us consider the barotropic equation of state $\kappa c^2 \rho = g(p_r)$. (2.6a)

On subtracting (2.3) from (2.5) gives

$$\left(\frac{\upsilon'+\lambda'}{r}\right)e^{-\lambda} = \kappa(c^2\rho + p_r) \tag{2.7a}$$

$$\left(\frac{\upsilon'+\lambda'}{r}\right)e^{-\lambda} = (g + \kappa p_r) \tag{2.7b}$$

Now, in order to solve (2.7b), let us further assume that metrics ($e^{\lambda}$ and $e^{\upsilon}$), and tangential pressure are arbitrary functions of radial pressure $p_r(\omega)$ such that $\omega$ is some function of $r$ i.e.

$$e^{-\lambda} = s(p_r(\omega)),\quad e^{\upsilon} = h(p_r(\omega)),\qquad \kappa c^2\rho = g(p_r(\omega)) \tag{2.6b}$$

Similar to approach taken by Bijalwan [1] substituting (2.6b) in (2.7b) leads to

$$\frac{(\bar{\upsilon} + \bar{\lambda})}{(c^2\rho + p)}e^{-\lambda}\frac{dp_r}{dr} = r \tag{2.8}$$

where overhead dash denotes derivative w.r.t. $p_r$ or $\omega$.

(2.8) yields

$$p_r = f(c_1 + c_2 r^2) = f(\omega)$$

i.e. function of '$c_1 + c_2 r^2$', where $c_1$ and $c_2$ ($\neq 0$) are arbitrary constants, such that

$$r = \sqrt{\frac{\omega - c_1}{c_2}},\ (\omega - c_1)c_2 > 0 \tag{2.9}$$

Assuming $f$ is invertible and $f^{-1}$ is inverse of $f$ then $f^{-1}(p) = \omega$.

Matter density, radial pressure, velocity of sound and tangential pressure can be expressed using (2.6) in (2.3), (2.4) and (2.5) as

$$c^2\rho = c_2\left(\frac{\bar{h}}{h}s - \bar{s}\right) - p_r(\omega) \tag{2.10}$$

$$p_r = c_2\left(2s\frac{\bar{h}}{h} - \frac{(1-s)}{(\omega - c_1)}\right) \tag{2.11a}$$

or $h = \exp\left(\int \frac{1}{s}\left(\frac{(1-s)}{(\omega-c_1)} + \frac{p_r}{c_2}\right)\right)$ (2.11b)

$$\sqrt{\frac{dp}{c^2 d\rho}} = 1/\sqrt{c_2\left(\overline{\left(\frac{\bar{h}}{h}s\right)} - \bar{\bar{s}}\right) - \bar{p}_r} \qquad (2.12)$$

$$\left[2\frac{\bar{h}}{h} + (\omega-c_1)\overline{\left(\frac{\bar{h}}{h}\right)} + (\omega-c_1)\left(\frac{\bar{s}}{s}\right)\left(\frac{\bar{h}}{h}\right) + (\omega-c_1)\left(\frac{\bar{h}}{h}\right)^2 + \frac{\bar{s}}{s}\right]s = \frac{p_\perp}{c_2} \qquad (2.15)$$

Equation (2.15) is first order linear differential equation for $s$ i.e.

$$\bar{s} + Zs = Z_1 \qquad (2.16)$$

where

$Z = \frac{\lfloor 2\alpha + f_1\bar{\alpha} + f_1\alpha^2 \rfloor}{[f_1\alpha + 1]}, Z_1 = \frac{\left[\frac{p_\perp}{c_2}\right]}{[f_1\alpha + 1]}$, such that $\alpha = \frac{\bar{h}}{h}$ and $f_1(p) = \omega - c_1$

Solving (2.16) leads to

$$s = e^{-\int z d\omega}\int e^{\int z d\omega} Z_1 d\omega \qquad (2.17)$$

The velocity of sound less than unity in the interior of star using (2.12) translates to

$$1 + \bar{p}_r < c_2\left(\overline{\left(\frac{\bar{h}}{h}s\right)} - \bar{\bar{s}}\right) \qquad (2.18)$$

Further, in order to have physically viable solutions, from (2.10) $c^2\rho + p \; (= c_2\left(\frac{\bar{h}}{h}s - \bar{s}\right))$ should be monotonically decreasing i.e.

$$2c_2^2\left(\overline{\left(\frac{\bar{h}}{h}s\right)} - \bar{\bar{s}}\right)r < 0 \qquad (2.19)$$

From (2.18) and (2.19) can only be satisfied if and only if $\left(\overline{\left(\frac{\bar{h}}{h}s\right)} - \bar{\bar{s}}\right) < 0$.

To illustrate, we choose $e^{-\lambda} = \omega^{-n}(c_3+\omega)^{-k}$ with $e^{\upsilon} = \omega^{n}(c_3+\omega)^{k}\exp(c_4(\omega - c_1))$ , Sharma [15]. From (2.11a) radial pressure would be

$$p_r = c_2\left(2(\omega^{-n}(c_3+\omega)^{-k})\left(\frac{n}{\omega} + \frac{k}{(c_3+\omega)} + c_4\right) - \frac{(1-\omega^{-n}(c_3+\omega)^{-k})}{(\omega-c_1)}\right) \qquad (2.11c)$$

We can always write $\alpha$ in terms of $p_r$ inverting (2.11c). Hence all physical parameters can be discussed in light of radial pressure.

**TABLE 1: METRICS FOR ANISOTROPIC FLUIDS**

| $e^{\upsilon}$ **[name [ref.]]** | $e^{-\lambda}\ (= e^{-\int z d\omega}\int e^{\int z d\omega} Z_1 d\omega)$ ( neutral or charged) <br> $f_1(p) = \omega - c_1,\ p_\perp = p_\perp(\omega)$ |
|---|---|
| $= h = c_3\begin{pmatrix} c_4\sqrt{1-\omega^2}\cos(c_4\sin^{-1}\omega) \\ +\omega\sin(c_4\sin^{-1}\omega)\end{pmatrix} + c_3\begin{pmatrix} c_4\sqrt{1-\omega^2}\sin(c_4\sin^{-1}\omega) \\ -\omega\cos(c_4\sin^{-1}\omega)\end{pmatrix}$ <br> ***[Mak [11]]*** | $Z = \dfrac{\lfloor 2\alpha + f_1\bar{\alpha} + f_1\alpha^2 \rfloor}{[f_1\alpha + 1]}, Z_1 = \dfrac{\left[\dfrac{p_\perp}{c_2}\right]}{[f_1\alpha + 1]}$ <br> $\alpha = \dfrac{\bar{h}}{h}$ |
| $= \omega^n (c_3 + \omega)^k \exp(c_4(\omega - c_1))$ <br> $= (c_1 + c_2 r^2)^n (c_3 + c_1 + c_2 r^2)^k \exp(c_4 c_2 r^2)$ <br> ***[Sharma[15]]*** | $Z = \dfrac{\lfloor 2\alpha + f_1\bar{\alpha} + f_1\alpha^2 \rfloor}{[f_1\alpha + 1]}, Z_1 = \dfrac{\left[\dfrac{p_\perp}{c_2}\right]}{[f_1\alpha + 1]}$ <br> $\alpha = \dfrac{n}{\omega} + \dfrac{k}{(c_3 + \omega)} + c_4$ |
| $= (c_3 - (\omega - c_1)^{(n+2)/2})^{-\frac{1}{(n+2)}}$ <br> $= (c_3 - c_2^{(n+2)/2} r^{n+2})^{-\frac{1}{(n+2)}}$ <br> ***[Rao [13]]*** | $Z = \dfrac{\lfloor 2\alpha + f_1\bar{\alpha} + f_1\alpha^2 \rfloor}{[f_1\alpha + 1]}, Z_1 = \dfrac{\left[\dfrac{p_\perp}{c_2}\right]}{[f_1\alpha + 1]}$ <br> $\alpha = \dfrac{(\omega - c_1)^{n/2}}{(c_3 + (\omega - c_1)^{(n+2)/2})}$ |
| $= \dfrac{c_3(\omega - c_1)^{\frac{1-m}{2m}}}{\left(c_4 - (\omega - c_1)^{\frac{(n+2)}{2}}\right)^{\frac{1}{m(n+2)}}}$ <br> $= \dfrac{c_3(c_2 r^2)^{\frac{1-m}{2m}}}{\left(c_4 - (c_2 r^2)^{\frac{(n+2)}{2}}\right)^{\frac{1}{m(n+2)}}}$ <br> ***[Rao [13]]*** | $Z = \dfrac{\lfloor 2\alpha + f_1\bar{\alpha} + f_1\alpha^2 \rfloor}{[f_1\alpha + 1]}, Z_1 = \dfrac{\left[\dfrac{p_\perp}{c_2}\right]}{[f_1\alpha + 1]}$ <br> $\alpha = \left(\dfrac{1-m}{2m}\right)\dfrac{c_3(\omega - c_1)^{\frac{1-3m}{2m}}}{\left(c_4 - (\omega - c_1)^{\frac{(n+2)}{2}}\right)^{\frac{1}{m(n+2)}}}$ <br> $+ \dfrac{c_3(\omega - c_1)^{\frac{1-m}{2m}+\frac{n}{2}}}{2m\left(c_4 - (\omega - c_1)^{\frac{(n+2)}{2}}\right)^{\frac{1}{m(n+2)}-1}}$ |

## 3. Solutions of Generalized TOV:

Now, we can specify two equations of state, such as $p_r = p_r(\rho)$ and $p_\perp = p_\perp(\rho)$, the equations (2.3)-(2.5) transform into a form where the hydrodynamical properties of the system are more evident. For systems with isotropic pressure, this formulation results in the Tolman-Oppenheimer-Volkov (TOV) equation. The generalized TOV equation, including anisotropy, is

$$\frac{dp_r}{dr} = -(\rho + p_r)\frac{\upsilon'}{2} + \frac{2}{r}(p_\perp - p_r) \qquad (3.1a)$$

with $\frac{\upsilon'}{2} = \frac{m(r) + 4\pi r^3 p_r}{r(r-2m)}$

and $m(r) = \int_0^r 4\pi r^2 \rho dr$ . (3.2a)

Taking $r = a$ ($a$ being radius of star) in the above expression gives us the Schwarzschild mass, $M = m(a)$ [This implicitly assumes that $\rho = 0$ for $r > a$.]

Using (2.6b) in (3.1a) and (3.2), we get

$$\overline{p_r} = \frac{1}{2c_2}\left(-(\rho + p_r)(\omega - c_1)\frac{\bar{h}}{h} + 2(p_\perp - p_r)\right) \qquad (3.1b)$$

$$m(r) = G/r \qquad (3.2b)$$

and $\rho = \frac{(2(\omega - c_1)\overline{G} - G)c_2^2}{4\pi(\omega - c_1)^2}$ (3.3)

where $G(\omega) = \frac{\left(\frac{\bar{h}}{h} - 4\pi\frac{p_r}{c_2}\right)(\omega - c_1)^2}{c_2\left(1 + 2(\omega - c_1)\frac{\bar{h}}{h}\right)}$ (3.4)

<u>Case I</u> : $\rho = \rho_0$, where $\rho_0$ is constant

Using (3.3), we have

$$\overline{G} - \frac{G}{2(\omega - c_1)} = \frac{2\pi(\omega - c_1)\rho_0}{c_2^2} \text{ or } G = \frac{4\pi(\omega - c_1)^2\rho_0}{3c_2^2}, \qquad (3.5)$$

$$\frac{\bar{h}}{h} = \frac{\frac{4\pi\rho_0}{3} + 4\pi p_r}{c_2\left(1 - \frac{8\pi(\omega - c_1)\rho_0}{3c_2}\right)}, \qquad (3.6)$$

and $\overline{p_r} = \frac{1}{2c_2}\left( 2(p_\perp - p_r) - 4\pi(\omega - c_1)\left( \frac{\frac{\rho_0^2}{3} + p_r^2 + \frac{4}{3}p_r\rho_0}{c_2\left(1 - \frac{8\pi(\omega - c_1)\rho_0}{3c_2}\right)} \right)\right)$ (3.7)

Dev [6] considered various forms of anisotropic factor

$$p_\perp - p_r = \frac{C}{c_2}(\omega - c_1)F(p_r, \rho_0)\left(1 - \frac{8\pi\rho_0}{3c_2}(\omega - c_1)\right)^{-1} \quad (3.1c)$$

for constant energy density $\rho_0$, where $C$ measures the amount of anisotropy and the function $F(p_r, \rho_0)$ which includes 6 separate cases,

$$F(p_r, \rho_0) = (p_r^2; p_r^2\rho_0; \rho_0^2; p_r^2 + p_r\rho_0; p_r^2 + \rho_0^2; p_r\rho_0 + \rho_0^2)$$

Bowers and Liang [2] solved the generalized TOV equation by considering the following equation of state,

$$p_\perp - p_r = \frac{C}{c_2}(\omega - c_1)\left(\frac{\rho_0^2}{3} + p_r^2 + \frac{4}{3}p_r\rho_0\right)\left(1 - \frac{8\pi\rho_0}{3c_2}(\omega - c_1)\right)^{-1} \quad (3.1d)$$

Let us consider here the a more general anisotropic factor

$$p_\perp - p_r = \frac{1}{2c_2}(\omega - c_1)\frac{\left(\frac{\rho_0^2}{3}(\beta_1 + 4\pi) + p_r^2(\beta_2 + 4\pi) + \frac{4}{3}p_r\rho_0(\beta_3 + 4\pi)\right)}{\left(1 - \frac{8\pi\rho_0}{3c_2}(\omega - c_1)\right)} \quad (3.1e)$$

where $\beta_i$'s are arbitrary constants, which yields

$$\frac{\overline{p_r}}{\left(\frac{\rho_0^2}{3}\beta_1 + p_r^2\beta_2 + \frac{4}{3}p_r\rho_0\beta_3\right)} = \frac{1}{2c_2^2}\frac{(\omega - c_1)}{\left(1 - \frac{8\pi\rho_0}{3c_2}(\omega - c_1)\right)} \quad (3.8)$$

(3.7) is easily integrable for different values of $\beta_i$'s. For brevity we don't discuss it in detail here.

<u>Case II</u> : $\rho$ is not constant

Let us explore a physically viable solution with NDG. On taking derivative of (3.3) w.r.t. $\omega$, we get

$$\overline{\rho} = \frac{(2(\omega - c_1)^2\overline{\overline{G}} - 3(\omega - c_1)\overline{G} + 2G)c_2^2}{4\pi(\omega - c_1)^3} \quad (3.9)$$

For $c_2 > 0$, NDG from (3.8) implies

$$2(\omega - c_1)^2 \overline{\overline{G}} - 3(\omega - c_1)\overline{G} + 2G < 0 \tag{3.10}$$

We consider here the anisotropy factor

$$p_\perp - p_r = \frac{C}{2}(\rho + p_r)(\omega - c_1)\frac{\overline{h}}{h}, \tag{3.1f}$$

where $C$ measures the amount of anisotropy.

From (3.1b) we have

$$\overline{p_r} = \frac{1}{2c_2}(C-1)(\rho + p_r)(\omega - c_1)\frac{\overline{h}}{h} \tag{3.11}$$

Let $G = \beta^2(\omega - c_1)^n$, $\beta$ and $n$ being arbitrary constants then (3.9) yields

$$(2n(n-1) - 3n + 2) < 0 \text{ or } \frac{1}{2} < n < 2 \tag{3.12}$$

Further, using (3.3) and (3.4) we have

$$\rho = \frac{(2n-1)\beta^2 c_2^2 (\omega - c_1)^{n-2}}{4\pi}, \tag{3.13}$$

$$\frac{\overline{h}}{h} = \frac{\left(c_2\beta^2(\omega - c_1)^{n-2} + 4\pi\frac{p_r}{c_2}\right)}{\left(1 - 2\beta^2 c_2(\omega - c_1)^{n-1}\right)}, \tag{3.14}$$

$$\overline{p_r} = \frac{1}{2c_2}(C-1)(\omega - c_1)\left(\frac{(2n-1)\beta^2 c_2^2(\omega - c_1)^{n-2}}{4\pi} + p_r\right)\frac{\left(c_2\beta^2(\omega - c_1)^{n-2} + 4\pi\frac{p_r}{c_2}\right)}{\left(1 - 2\beta^2 c_2(\omega - c_1)^{n-1}\right)} \tag{3.15}$$

We may here take $p_r = \beta_1^2(\omega - c_1)^m$, $\beta_1$ and $m$ being arbitrary constants to satisfy (3.15) yielding

$$\beta_1^2 m(\omega - c_1)^{m-2}\left(1 - 2\beta^2 c_2(\omega - c_1)^{n-1}\right)$$
$$= \frac{1}{2c_2}(C-1)\left(\frac{(2n-1)\beta^2 c_2^2}{4\pi} + \beta_1^2(\omega - c_1)^{m-n+2}\right)\left(c_2\beta^2 + 4\pi\frac{\beta_1^2(\omega - c_1)^{m-n+2}}{c_2}\right) \tag{3.16}$$

Above inequality can only be satisfied for $m = 2$ which implies positive radial pressure gradient. Hence, these solutions are suitable for modeling expanding anisotropic fluids. In order to explore special cases of solutions by Chaisi [4], Dev [6] and others, we can choose

$$G = \frac{4\pi j(\omega - c_1)}{c_2} + \frac{4\pi k}{3c_2^2}(\omega - c_1)^2 + \frac{4\pi l}{c_2^3}(\omega - c_1)^3 \tag{3.17}$$

where $l$, $j$ and $k$ are arbitrary constant.

Using (3.10) and (3.17), we have

$$2(\omega-c_1)^2\left(\frac{8\pi k}{3c_2^2}+\frac{24\pi l}{5c_2^3}(\omega-c_1)\right)-3(\omega-c_1)\left(\frac{4\pi j}{c_2}+\frac{8\pi k}{3c_2^2}(\omega-c_1)+\frac{12\pi l}{5c_2^3}(\omega-c_1)^2\right)$$
$$+2\left(\frac{4\pi j(\omega-c_1)}{c_2}+\frac{4\pi k}{3c_2^2}(\omega-c_1)^2+\frac{4\pi l}{5c_2^3}(\omega-c_1)^3\right)<0 \tag{3.18}$$

which leads to $l<\dfrac{jc_2^2}{(\omega-c_1)^2}$. (3.19)

Using (3.3) and (3.4), we get

$$\rho=\frac{jc_2}{(\omega-c_1)}+k+l\frac{(\omega-c_1)}{c_2} \tag{3.20}$$

$$\frac{\bar{h}}{h}=\frac{\left(4\pi j(\omega-c_1)+\dfrac{4\pi k}{3c_2}(\omega-c_1)^2+\dfrac{4\pi l}{c_2^2}(\omega-c_1)^3+4\pi\dfrac{p_r}{c_2}(\omega-c_1)^2\right)}{\left((\omega-c_1)^2-8\pi j-\dfrac{8\pi k}{3c_2}(\omega-c_1)-\dfrac{8\pi l}{c_2^2}(\omega-c_1)^2\right)} \tag{3.21}$$

As our goal is to present a new methodology we don't investigate here any other solutions with NDG.

## 4. Conclusions

We have studied the structure of the sources produced by Einstein field equations with an anisotropic matter distribution in terms of pressure for a generalized TOV equation. Further, we present a new methodology to derive solutions with NDG.